# Helicoidal magnetic ordering in crystals: exact periodic solutions of equations of state with fourth-order anisotropy.

Alexander Ya. Braginsky

Southern Federal University, Physics Research Institute, Rostov-on-Don

e-mail: **a.braginsky@mail.ru**

**Abstract.**

In this paper we obtain an exact periodic solution for the system of equations of state corresponding to the helicoidal phase of crystalline ordering. We used Gufan's method of integral rational basis of invariants (IRBI) to construct an inhomogeneous Landau potential. As a result the inhomogeneous Landau potential takes account of anisotropic invariants consisting of OP components as well as anisotropic invariants that comprise space derivatives of OP components. It is demonstrated that it is taking into account the latter that leads to the exact periodic solution for the system of equations of state that describes the helicoidal phase. A phase diagram of states is built for active Lifshitz representations at the second-order phase transition point. Analogy between inhomogeneous states of helimagnetics and second-order superconductors in the magnetic field is discussed. It is suggested to describe the vortex state of a helimagnetic in the magnetic field as a deformation of the magnetic sublattice with dislocations and not as the state of skyrmions.

## 1. Introduction

As is known[1-3], long-period helicoidal structures do exist in crystals. The evidence for this is provided in neutron diffraction studies of magnetic substances [4] that have shown that a number of crystals have phase transitions with the formation of spiral structures. They are characterized by periodic dependence of the magnetic moment (for example) on the coordinate. However, description of helicoidal structures in crystals is not covered in the inhomogeneous Landau theory because the system of equations of state studied up until now did not allow for exact periodic solutions. The exact periodic solution was obtained for inhomogeneous Landau potential at approximation of the absence of anisotropy [5]. In this case, essentially, equation $\delta\Phi/\delta\varphi = 0$ was neglected, where $\Phi$ is the thermodynamic potential and $\varphi$ is the OP phase. However, in the absence of anisotropy equations of state describe isotropic continuum which does not agree with the description of the crystal. Also periodic solutions were obtained in approximation $\rho = const$ in the form of elliptic functions [6,7], where $\rho$ is the OP module. However, this approximation neglected one of the equations of the system of equations of state, namely $\delta\Phi/\delta\rho = 0$. Such an approximation is probably acceptable in a description away from the phase transition point but it is not acceptable when describing the symmetry of state in the neighbourhood of the second-order phase transition point. As is known, when studying the second-order phase transition at a certain temperature a solution appears in which the OP module is non-zero.

As there is no exact periodic solution for the full system of equations of state, it was suggested that the helicoid symmetry should be characterised by the first harmonic expansion of an inhomogeneous solution into the Fourier series. However, such a suggestion does not stand up to scrutiny because any continuously differentiable function can be expanded into a Fourier series but its translational symmetry is not defined by the first harmonic. Symmetry is an exact concept, therefore, approximate solutions or solutions for an incomplete system of equations of state are not suitable for the classification of physical states by symmetry.

Thus, long-period helicoidal structures can be seen in crystals [1-4] but they are not described by the existing theory of the phase transitions into inhomogeneous states [5,8]. This is, obviously, a gap in the theory. This paper will demonstrate that the matter is in the method used to construct the Landau potential to describe inhomogeneous states. The problem is eliminated when the Landau potential takes into account all the anisotropic invariants of OP components and their derivatives that are determined by the symmetry of the crystal.

The nonequilibrium Landau potential is a function of the independent invariants of the order parameter components. To construct the Landau potential the method of the integral rational basis of invariants (IRBI) was developed [9]. This allowed the construction of phase diagrams that were not described by Landau potentials without taking into account all the invariants included the IRBI [10].

In this paper we will use Gufan's method and will construct the IRBI not only for OP components but for their space derivatives as well. Let us remember, however, that in variational calculus the OP components and their space derivatives are considered to be independent degrees of freedom when constructing the nonequilibrium potential. Then the inhomogeneous Landau potential will contain anisotropic invariants made of not only OP components but of their space derivatives as well. It should be noted that space derivatives of OP components usually make part of the Landau potential in the form of quadratic invariants, for example $\eta_{1z}^2 + \eta_{2z}^2$, or antisymmetric Lifshitz invariants, for example $\eta_1 \eta_{2z} - \eta_2 \eta_{1z}$. It will be demonstrated later that taking into account anisotropic invariants that contain space derivatives of OP, for example, $\eta_{1z}^2 \eta_{2z}^2$, results in an exact periodic solution of the system of equations of state that corresponds to the helicoidal phase.

## 2. Inhomogeneous Lifshitz model.

The inhomogeneous theory of phase transitions with OP depending on macroscopic coordinates is based on a formalism constructed by Lifshitz [8]. Generally speaking, OP cannot depend on a coordinate by definition because OP are coefficients of state density expansion in irreducible representation (IR) of the crystal space group. As is known [], the translational symmetry of the crystal is determined by the crystal lattice and the IR of the translations subgroup are equivalent to set $\exp(i\vec{k}_l \vec{x})$. So expansion of the state density in irreducible representations (IR) of the crystal space group is a generalisation of the Fourier series expansion of a function. It is obvious that coefficients of the Fourier series expansion for a function do not depend on the coordinate. That is why before we introduce the dependence of OP on the coordinate it needs to be clarified what we mean by this dependence. In paper [8] Lifshitz looked into spatial fluctuations of OP and proved the inhomogeneous model within the phenomenological Landau theory. In his formalism Lifshitz believed that:

1) A crystal can be considered homogeneous in a macroscopically small volume.

2) The macroscopically small volume of a crystal is characterised by a macroscopic coordinate $\vec{X}$.

3) The transition from a high-symmetry phase to a low-symmetry phase occurs as per one IR of the space group in each point $\vec{X}$.

4) Expansion coefficients of density of state in IR basic functions depend on macroscopic coordinates $\eta_l = \eta_l(\vec{X})$.

5) Local nonequilibrium thermodynamic potential $\Phi(\vec{X})$ is functional $\eta_l(\vec{X})$ and depends invariantly on order parameter (OP) components and its space derivatives: $\eta_l(\vec{X})$, $\eta_{l,X_j}(\vec{X})$, (here designation $\eta_{l,X_j}(\vec{X}) \equiv \partial \eta_l / \partial X_j$).

6) Equilibrium distributions of OP in the crystal correspond to extremals of functional $\Phi(\vec{X})$.

Dzyaloshinskii [5] used the inhomogeneous Lifshitz formalism to describe phase transitions in which a crystal goes from a state with relatively small primitive period to a state without any translational symmetry. He demonstrated that if the local nonequilibrium thermodynamic potential contains a Lifshitz invariant (antisymmetric quadratic combination, linear in OP components as well as in its gradients) then the high-symmetry phase goes into an inhomogeneous phase through second-order transition at temperature $T_0$ above Curie point $T_c$ of the possible phase transition into a low-symmetry homogeneous phase. It was suggested that the resulting inhomogeneous structure would be a helicoidal phase that corresponds to an exact periodic solution of the Euler-Lagrange system of equations for $\Phi(\vec{X})$. However, the exact periodic solution was not obtained in the Dzyaloshinskii's model [5]. The presence of the fourth-order anisotropic invariant that characterises the symmetry of the orthorhombic crystal was incompatible with equation of state $\delta\Phi/\delta\varphi = 0$, where $\varphi$ is an OP phase for the sinusoidal solution. Further analysis of the Euler-Lagrange system of equations in the Dzyaloshinskii's model [5] showed that its solutions are inhomogeneous and contain, in the general case, an infinite number of harmonics. The attempt to describe helicoidal structures through approximated solutions where the period is determined by the first harmonic of the Fourier series expansion of an inhomogeneous solution does not stand up to scrutiny. In essence, the model of the potential studied in [5], could not provide description for observed helicoidal structures in crystals.

In this paper we obtain an exact periodic solution for the system of equations of state taking full account of anisotropy, the solution that corresponds to helicoidal ordering in crystals (Chapter 3). It is proven that solutions to the Euler-Lagrange system of equations where the OP module is dependent on space coordinates will satisfy inhomogeneous states different from helicoidal states (Chapter 4). We offer a description for magnetic vortex structures in the neighbourhood of the second-order phase transition (Chapter 5). We study the diagram of states in the neighbourhood of the point of the second-order phase transition from the high-symmetry phase to the low-symmetry phase when the density of inhomogeneous potential $\Phi(\vec{X})$ contains Lifshits invariant (Chapter 6). In Chapter 7 we trace an analogy in the description of inhomogeneous helimagnetics [11] and second-order superconductors. The vortex state of helimagnetics is described as a deformation of the magnetic sublattice with the occurrence of dislocations, similar to Abrikosov vortexes [12].

**3. Exact periodic solution for a two-component OP.**

As an example, let us examine structural transitions in an orthorhombic crystal with a symmetry group $D_{2h}^{16}$. Let the ordering of the low-symmetry phase be described with the OP that is transformed in IR $\tau_2$ of star $\vec{k}_{21}$ [13]. We chose this representation because it allows for the Lifshitz invariant, an antisymmetry combination of OP components and its derivatives. Such a representation describes antiferromagnetic ordering phase transitions in an orthorhombic crystal and phase transitions to inhomogeneous states.

To construct local Landau potential let us use the IRBI method [10] that allows to take into account all the anisotropic interactions of OP components that are determined by the symmetry of the crystal. Let us find the matrices for the representation of the space group of symmetry of the crystal in order to construct the IRBI of OP components and its space derivatives. They act in a 4-D space of OP components and their derivatives: $\eta_1$, $\eta_2$, $\eta_{1z}$, $\eta_{2z}$; where $Oz$ is the axis of symmetry. To make it simple, we will consider only one-parameter dependence of OP components on the coordinate and we will designate it with a small letter $z \equiv X_3$, bearing in mind all along that this refers to the macroscopic Lifshitz coordinate. The IRBI for this representation has the following form:

$$I_1 = \eta_1^2 + \eta_2^2, \quad I_2 = \eta_1\eta_{2z} - \eta_2\eta_{1z}, \quad I_3 = \eta_{1z}^2 + \eta_{2z}^2, \quad I_4 = \eta_1^2\eta_2^2, \quad I_5 = \eta_1\eta_2\eta_{1z}\eta_{2z},$$
$$I_6 = \eta_1\eta_{2z}^3 - \eta_2\eta_{1z}^3, \quad I_7 = \eta_{1z}^2\eta_{2z}^2, \quad I_8 = \eta_1^3\eta_{2z} - \eta_2^3\eta_{1z}. \tag{1}$$

$I_1, I_2, I_3$ are isotropic invariants, and - $I_4, I_5, I_6, I_7, I_8$ are anisotropic invariants. The representation we chose allows for the Lifshitz invariant - $I_2$, therefore, inhomogeneous states in the neighbourhood of the second-order phase transition point will possess minimal energy. In this case it is necessary to investigate local Landau potential $\Phi(\vec{X})$ that is a functional of OP components and their space derivatives.

Before we proceed to constructing $\Phi(\vec{X})$, it must be noted that:

$$\eta_1^3\eta_{2z} - \eta_2^3\eta_{1z} = \frac{1}{4}\left[3(\eta_1^2 + \eta_2^2)(\eta_1\eta_{2z} - \eta_2\eta_{1z}) + \frac{d}{dz}(\eta_1^3\eta_2 - \eta_2^3\eta_1)\right].$$

This leads to a conclusion that $I_8 = \frac{3}{4}I_1 I_2$ plus a full derivative of expression $\eta_1^3\eta_2 - \eta_2^3\eta_1$ that, when integrated, provides constant contribution to the thermodynamic potential insignificant to determine the minimum functional. Therefore, instead of invariant $I_8$ it will suffice to take into account the product of invariants $I_1 I_2$. Then functional $\Phi(\vec{X})$ of the fourth degree in OP components and its space derivatives that takes into account all invariants (1) will have the form :

$$\Phi = a_1 I_1 + a_2 I_2 + a_3 I_3 + a_{11} I_1^2 + a_{12} I_1 I_2 + a_{13} I_1 I_3 + a_{22} I_2^2 + a_{23} I_2 I_3 + a_{33} I_3^2 +$$
$$+ b_1 I_4 + b_2 I_5 + b_3 I_6 + b_4 I_7. \tag{2}$$

As the Euler-Lagrange equations:

$$\frac{\partial \Phi}{\partial \eta_1} - \frac{\partial}{\partial z}\left(\frac{\partial \Phi}{\partial \eta_{1z}}\right) = 0, \quad \frac{\partial \Phi}{\partial \eta_2} - \frac{\partial}{\partial z}\left(\frac{\partial \Phi}{\partial \eta_{2z}}\right) = 0 \tag{3}$$

can be solved in relation to higher derivatives, then in the general case each boundary condition is satisfied with a well defined distribution of OP throughout the crystal. For the inhomogeneous problem the range of solutions for minimisation equations is continuous (as opposed to the homogeneous problem where the range is discrete), so the main state for fixed thermodynamic parameters is selected by way of additional minimisation of nonequilibrium thermodynamic potential as per boundary conditions.

In order to classify solutions to Euler-Lagrange equations by symmetry let us move to the spherical coordinate system of OP, $\eta_1 = \rho \cos\varphi$, $\eta_2 = \rho \sin\varphi$. In this case the homogeneous phase is satisfied with a specific solution for the system of second-order differential equations $\rho_c = const$, $\varphi_c = const$ with boundary conditions $\rho|_B = \rho_c$, $\varphi|_B = \varphi_c$, corresponding to minima of the homogeneous Landau potential. Similarly, the helicoidal phase is satisfied with a specific solution: $\rho_h = const$, $\varphi_h = q_h z$ where $q_h = const$ with boundary conditions $\rho|_B = \rho_h$, $\partial\varphi/\partial z|_B = q_h$. Boundary conditions in the above examples are obtained by way of substitution of corresponding class of solutions in the system of equations of state, and , $\rho_c$, $\varphi_c$, $\rho_h$, $q_h$ are defined in it as functions from coefficients of the potential that depend, in their turn, on thermodynamic conditions set on the thermostat.

Let us look for a solution for system (3) in this form:
$$\eta_1 = \rho \cos qz, \quad \eta_2 = \rho \sin qz, \tag{4}$$
where $\rho = const$, $q = const$. Here we introduced designation $q$ on purpose for the period of the helicoid in order not to confuse it with IR vector $\vec{k}$ that is set when choosing the OP (generally speaking, these are different objects; $q$ characterizes inhomogeneity in the macrospace, and vector $\vec{k}$ characterizes transformational properties of the OP in translation to the primitive period). By multiplying the first equation (3) by $\eta_1$ and the second equation by $\eta_2$ and adding them together and inserting (4) we obtain the first equation of the system (5). By multiplying the first equation (3) by $\eta_2$ and the second equation by $\eta_1$ and subtracting the second equation from the first equation, bearing (4) in mind, we obtain the second equation of the system (5):

$$2(a_1 + a_2 q + a_3 q^2)\rho^2 + [(4a_{11} + \tfrac{1}{2}b_1) + 4a_{12}q + (4a_{22} + 4a_{13} - \tfrac{1}{2}b_2)q^2 +$$
$$+ (4a_{23} + 3b_3)q^3 + (4a_{33} + \tfrac{1}{2}b_4)q^4]\rho^4 - (\tfrac{1}{2}b_1 + \tfrac{1}{2}b_2 q^2 + 2b_3 q^3 - \tfrac{3}{2}b_4 q^4)\rho^4 \cos 4qz = 0$$
$$\tag{5}$$

$$(\tfrac{1}{2}b_1 + \tfrac{1}{2}b_2 q^2 + 2b_3 q^3 - \tfrac{3}{2}b_4 q^4)\rho^4 \sin 4qz = 0.$$

System of equations (5) comes down to an algebraic system of two equations in relation to $\rho$ and $q$:

$$[(2a_{11} + \tfrac{1}{4}b_1) + 2a_{12}q + (2a_{22} + 2a_{13} - \tfrac{1}{4}b_2)q^2 + (2a_{23} + \tfrac{3}{2}b_3)q^3 +$$
$$+ (2a_{33} + \tfrac{1}{4}b_4)q^4]\rho^2 = -(a_1 + a_2 q + a_3 q^2)$$
$$\tag{6}$$

$$b_1 + b_2 q^2 + 4b_3 q^3 - 3b_4 q^4 = 0.$$

The solution for system (6) exists when $\rho^2 \geq 0$, which is equal to condition $a_1 + a_2 q + a_3 q^2 \leq 0$. The multiplier preceding $\rho^2$ in the first equation should be positive for the reasons of thermodynamic stability of the periodic solution. According to (6), the period of helicoid is defined only by anisotropic coefficients of the potential in the second equation.

Substituting solutions $q$ from the second equation of system (6) in the first equation we obtain an expression for $\rho$.

It should be noted that in case of absence of anisotropy when all $b_i = 0$, $q$ and $\rho$ are connected only in one equation (6). Moreover, if counting only quadratic invariants in the potential, the system of equations of state for an isotropic case is reduced to the first part of the first equation of the system (6) being equal to zero: $a_1 + a_2 q + a_3 q^2 = 0$. Such models are usually considered in order to determine physical quantities in the neighbourhood of the second-order phase transition when OP values are small. For example, quadratic approximation is used in skyrmion theory []. But it follows from expression $a_1 + a_2 q + a_3 q^2 = 0$ that in linear approximation helicoid period $q$ should only be a function of isotropic coefficients of the potential. This does not agree with the conclusion derived from the system of Euler-Lagrange equations of state (3) that helicoid period $q$ depends only upon anisotropic coefficients of the potential (second equation of system (6)). We need to eliminate this contradiction.

For potential (2) condition $a_1 + a_2 q + a_3 q^2 = 0$ is equivalent to the coefficient preceding $\rho^2$ being equal to zero; this can be easily seen if (4) is inserted in (2). Such a condition is not dictated by the symmetry. Approximation of quadratic potential for OP is incorrect in principle as per the non-linear Landau theory. Let us demonstrate that such an approximation is also incorrect in a particular case for solution (4) that corresponds to helicoidal ordering in the crystal. Indeed, linear approximation $a_1 + a_2 q + a_3 q^2 = 0$ for sinusoidal solution (4), in the general case, is not compatible with expression $b_1 + b_2 q^2 + 4 b_3 q^3 - 3 b_4 q^4 = 0$ that is derived from the system of Euler-Lagrange equations of state. Moreover, for helicoidal solution (4) expression $a_1 + a_2 q + a_3 q^2 \neq 0$, as $\rho^2 \neq 0$ from the first equation (6). Essentially, condition $a_1 + a_2 q + a_3 q^2 = 0$ is a degeneracy condition for the helicoidal solution because if $a_1 + a_2 q + a_3 q^2 = 0$ in the Landau theory $\rho = 0$ (5, 6). It should be noted that when $\rho = 0$ all OP components are also equal to zero (4) and there is no helicoidal solution. Therefore, helicoid period $q$ in the crystal is only given by anisotropic coefficients of the potential (second equation (6)), and $\rho$ is defined by isotropic as well as by anisotropic coefficients of the potential from the first equation in system (6). Thus, linear approximation (quadratic approximation of the potential) must not be used when describing helicoidal states in the Landau theory.

The Dzyaloshinskii's model [5] did not take into account anisotropic gradient basic invariants of this form $-I_5, I_6, I_7$ case $b_2 = b_3 = b_4 = 0$ in (2, 6). Therefore, equation for minimisation in the OP angular coordinate $\delta\Phi/\delta\varphi = 0$ for the periodic solution equivalent to the second equation of system (6) would look like this: $b_1 = 0$. This means that the equation of state would not be an algebraic expression for parameters of sinusoidal solution (4) $q$ and $\rho$, but a condition for being equal to zero for anisotropic coefficient of the potential. This condition means that in the Dzyaloshinskii's model the system of equations of state is incompatible with the sinusoidal solution that corresponds to helicoidal phase of the crystal when $b_1 \neq 0$. This led to a method of obtaining a periodic solution "in approximation of neglect with anisotropy" when

$b_1 = 0$ was assumed. As a matter of fact, it came down to neglecting equation $\delta\Phi/\delta\varphi = 0$ that corresponds to the second equation of system (6). The case when $b_1 = 0$ and $b_2 = b_3 = b_4 = 0$, that is to say when all anisotropic coefficients of potential (2) are equal to zero, satisfies the approximation of the isotropic medium. It is obvious that such an approximation is not suitable for the description of ordering in orthorhombic crystals.

## 4. Analysis of solutions for the system of equations of state with $\rho = const$.

Due to the lack of an exact periodic solution with condition $b_1 \neq 0$, paper [7] put forward an assumption that the OP module is not dependent on coordinates in the neighbourhood of the point of second-order phase transition from a high-symmetry phase to an inhomogeneous phase. It was suggested that helicoidal structure should be described using the Dzyaloshinskii's elliptic periodic solution [6] obtained to describe periodic structure away from the second-order phase transition point. Additionally in [6] it was assumed that $\rho = const$ and that one of the equations in the system of equations of state can be neglected, namely $\delta\Phi/\delta\rho = 0$. Fixing the degree of freedom in variational calculus corresponds to a problem with condition and can be solved through the introduction of an additional Legendre factor $\lambda = \delta\Phi/\delta\rho$, which was done in [6]. The assumption that the degree of freedom is not dependent on coordinates in the neighbourhood of the second-order phase transition point is not equivalent to the assumption about fixing the degree of freedom. Searching for a solution in the form $\rho = const$ and solving a system of equations of state with condition $\rho = const$ are two separate problems. In the first case we insert $\rho = const$ in the full system of equations of state and look for parameter $\rho$. In the second case it is assumed that solution $\rho = const$ exists (is fixed) and then a problem with a condition is being solved, without taking into account equation $\delta\Phi/\delta\rho = 0$ in the system of equations of state. It is obvious that in the neighbourhood of the second-order phase transition we cannot talk of fixing the module of OP that defined the phase transition. Therefore, value $\rho$ should be determined from the full system of equations of state. Let us demonstrate that when $\rho = const$ the system of equations of state (3) does not have solutions different from $\varphi_z = const$. Here and in what follows, we will understand $\rho = const$ as independence of $\rho$ from the coordinate, namely $\rho_z = 0$.

Let us look for solutions to the system of equations of state (3) in the form $\eta_1 = \rho\cos\varphi$, $\eta_2 = \rho\sin\varphi$, where $\rho = const$, $\varphi = \varphi(z)$. We obtain:

$$2(a_1 + a_2\varphi_z + a_3\varphi_z^2)\rho^2 + [4a_{11} + \tfrac{1}{2}b_1 + 4a_{12}\varphi_z + (4a_{22} + 4a_{13} - \tfrac{1}{2}b_2)\varphi_z^2 +$$
$$(4a_{23} + 3b_3)\varphi_z^3 + (4a_{33} + \tfrac{1}{2}b_4)\varphi_z^4]\rho^4 - (\tfrac{1}{2}b_1 + \tfrac{1}{2}b_2\varphi_z^2 + 2b_3\varphi_z^3 - \tfrac{3}{2}b_4\varphi_z^4)\rho^4\cos4\varphi -$$
$$-(\tfrac{1}{4}b_2 + \tfrac{3}{2}b_3\varphi_z - \tfrac{3}{2}b_4\varphi_z^2)\varphi_{zz}\rho^4\sin4\varphi = 0$$

(7)

$$[2a_3 + (2a_{13} + 2a_{22} - \tfrac{1}{4}b_2)\rho^2 + (6a_{23} + \tfrac{9}{2}b_3)\rho^2\varphi_z + (12a_{33} + \tfrac{3}{2}b_4)\rho^2\varphi_z^2 +$$
$$+(\tfrac{1}{4}b_2 + \tfrac{3}{2}b_3\varphi_z - \tfrac{3}{2}b_4\varphi_z^2)\rho^2\cos4\varphi]\rho^2\varphi_{zz} -$$
$$-(\tfrac{1}{2}b_1 + \tfrac{1}{2}b_2\varphi_z^2 + 2b_3\varphi_z^3 - \tfrac{3}{2}b_4\varphi_z^4)\rho^4\sin4\varphi = 0$$

It should be noted that the system of equations (7) can be solved with respect to $\rho$. By multiplying the second equation by $(\frac{1}{4}b_2 + \frac{3}{2}b_3\varphi_z - \frac{3}{2}b_4\varphi_z^2)\rho^4 \sin 4\varphi$ and the first equation by the factor preceding $\varphi_{zz}$ in the second equation and by adding them together we obtain algebraic equation:

$$P_1(\cos 4\varphi, \varphi_z, \rho^2) = 0 \qquad (8)$$

We obtain the second polynomial $P_2(\cos 4\varphi, \varphi_z, \rho^2) = 0$ from the first integral for potential $\Phi(\vec{X})$ that, as per Noether's theorem, has the form:

$$\frac{\partial \Phi}{\partial \eta_{1z}} \eta_{1z} + \frac{\partial \Phi}{\partial \eta_{2z}} \eta_{2z} - \Phi = const \qquad (9)$$

Expression (9) can be obtained by way of integrating (3) with integrating multipliers. Indeed, let us multiply the first equation by $\eta_{1z}$, the second equation by $\eta_{2z}$, then add them together and obtain the total differential: $\frac{d\Phi}{dz} - \frac{d}{dz}\left(\frac{\partial \Phi}{\partial \eta_{1z}} \eta_{1z} + \frac{\partial \Phi}{\partial \eta_{2z}} \eta_{2z}\right) = 0$. After integrating we have (9). Expression (9) also has the form $P_2(\cos 4\varphi, \varphi_z, \rho^2) = 0$. This can be easily seen by inserting invariants (1) into (2, 9) in spherical coordinates, $\eta_1 = \rho \cos \varphi$, $\eta_2 = \rho \sin \varphi$ provided that $\rho = const$. It should be noted that polynomial $P_2(\cos 4\varphi, \varphi_z, \rho^2)$ does not depend on $\sin 4\varphi$, as $\sin 4\varphi$ only makes part of anisotropic invariants $I_5, I_6, I_7$ (1) in combination with factor $\rho_z = \partial \rho / \partial z$, and we assume that $\rho_z = 0$.

Thus, we have moved from system (7) to the equivalent system (8, 9) of algebraic equations for, $\cos 4\varphi$, $\varphi_z$ and $\rho^2$. In the general case such system of equations has a solution provided that resultant

$$Res(P_1(\cos 4\varphi), P_2(\cos 4\varphi)) = 0 \qquad (10)$$

is equal to 0. Then the resultant itself is a polynomial for $\varphi_z$ and $\rho^2$: $Res = P_3(\varphi_z, \rho^2)$. Equation $P_3(\varphi_z, \rho^2) = 0$ is satisfied for any $z$ only provided $\varphi_z = const$, q.e.d.

If $\varphi_z = 0$, then all possible homogeneous states are listed from the second equation of system (7), $\sin 4\varphi = 0$ and from the solution for system (7). Case $\varphi_z = const \neq 0$ was examined in Chapter 3 and it satisfies the helicoidal phase.

Similarly it is possible to demonstrate that system (3) does not have solutions in the form $\rho = \rho(z)$, $\varphi = qz$, where $q = const$.

Thus, the system of Euler-Lagrange equations (7) does not have periodic solutions like elliptic functions [7, 8]. It follows from the presented analysis of the system of equations (3) that inhomogeneous state other than helicoidal state is satisfied with essentially inhomogeneous solution where $\rho = \rho(z)$, $\partial \varphi / \partial z = f(z)$.

If quadratic approximation of the potential is used to determine $\varphi(z)$, then the obtained solutions will result in the coefficient preceding $\rho^2$ being equal to zero in the first equation of system (7). This leads to a conclusion that $\rho = 0$, and so all OP components are equal to zero for $\varphi(z)$ obtained from linear approximation of equations of state. Therefore, quadratic approximation of the potential must not be used in Landau theory for calculations of the OP phase when $\rho = const$. If the OP depend on $\vec{X}$, the reasoning will not change significantly.

**5. Magnetic vortex states.**

In recent years, inhomogeneous magnetic structures in crystals were studied within the theory of magnetic vortex structures - chiral skyrmions [14,15]. In papers [15], the nonequilibrium potential takes into account invariant $\vec{M}[\vec{\nabla}\vec{M}]$ that generalizes the Lifshitz invariant for the case when several coordinates are involved in an antisymmetry invariant combination. Such invariants exist for crystals that do not have an inversion centre. When constructing the inhomogeneous Landau potential in [15] it is assumed that $\vec{M}(\vec{X})$ represents a vector field, therefore, it is transformed in IR with $\vec{k} = 0$. It is believed that such invariants describe the Dzyaloshinskii-Moriya interaction $[\vec{S}_i \vec{S}_j]$ [16] in the phenomenological Landau theory. In calculations in [15] $|\vec{M}| = const$ is fixed out of additional considerations and the nonequilibrium potential is minimised by $\vec{n}(\vec{X}) = \vec{M}/|\vec{M}|$. This means that in [14,15] the problem is being solved with the condition $|\vec{M}| = const$.

However, the authors of [15,16] limit their calculations to quadratic potential. Let us demonstrate that the reasoning given above is also valid for crystals without an inversion centre. As is known, approximation of quadratic potential is justified in the small neighbourhood of the point of the second-order phase transition when the OP is small and its higher degrees can be neglected in the nonequilibrium potential. On the other hand, in the neighbourhood of the second-order phase transition point non-linear equation $\delta \Phi / \delta |\vec{M}| = 0$ must be taken into account because when a new phase appears it is characterised by solutions of a non-linear equations of state for which $|\vec{M}| \triangleright 0$. Therefore, in the small neighbourhood of the second-order phase transition point [] a full system of equations of state needs to be solved that would contain, among others, $\delta \Phi / \delta |\vec{M}| = 0$.

If $|\vec{M}| = const$, then equation $\delta \Phi / \delta |\vec{M}| = 0$ can be solved for $|\vec{M}|$. In this case, as was demonstrated above, quadratic approximation of the potential for $\vec{n}(\vec{X}) = \vec{M}/|\vec{M}|$ leads to solution $|\vec{M}| = 0$ for non-linear equation $\delta \Phi / \delta |\vec{M}| = 0$.

Nonetheless, in a number of cases [15] solutions for a variations problem with condition $|\vec{M}| = const$ in quadratic approximation provide a quality description for inhomogeneous vortex distribution of the magnetic moment [11]. There is also an alternative point of view to describe these structures [17]. In our opinion, from the very beginning [5] the problem of the long-period structures was associated with equating of the helicoid period $\vec{q}$ (4) and IR vector $\vec{k}$. Now, when we obtained the exact periodic solution (6) that corresponds to the helicoidal phase (4), we can consider the IR where vector $\vec{k}$ does not initially correspond to the chosen point of the Brillouin zone. Such an IR is not equivalent to the vector IR or IR with vector $\vec{k}_0$, that correspond to the chosen point of the Brillouin zone just for the simple fact that they have different number of OP components and they transform differently from the crystal symmetry group.

As an alternative for skyrmions another description of magnetic vortex states can be suggested [18] where the OP with $\vec{k} \neq 0$ is used as the main variable. It is determined by the translational symmetry of the potential and, as a consequence, by the minimal interaction between the magnetic OP and distortion tensor. In this case the vortex structure of equations of state will be related to gradient invariance inherent to equations that contain distortion tensor, similar to Maxwell equations.

Dislocations as functions of state appear in the inhomogeneous Landau theory if Lifshitz's approach formulated in Chapter 2 is generalised for a case when not only the OP value changes with $\vec{X}$ but also the OP transformational properties with respect to the translation subgroup change with the macro-coordinate. Indeed, in translations to lattice spacing $\vec{a}$ in the general case the OP transformational properties are characterised by a continuous parameter - IR vector $\vec{k}$: $\hat{\vec{a}} \eta_l = \exp(i\vec{k}_l \vec{a}) \eta_l$, that can also change with the macro-coordinate $\vec{k}_l = \vec{k}_l(\vec{X})$. Here, one should not confuse IR vector $\vec{k}$ with helicoid period $\vec{q}$ that characterizes the value of OP components (4). In the case when $\vec{k}_l = \vec{k}_l(\vec{X})$ under the effect of translation operator $\hat{\vec{a}}$ value $i\eta_l \partial(\vec{k}_l \vec{a})/\partial X_j$ will be added to derivative $\partial \eta_l / \partial X_j$, that value needs to be annulled in order to construct the translation-invariant potential. In order to annul gradient $i\partial(\vec{k}_l \vec{a})/\partial X_j$ let us introduce an additional compensating field to the extended derivative of the OP. We will construct the extended derivative by analogy with the field theory of electrodynamics, bearing in mind that here the local group parameter will be IR vector $\vec{k}(\vec{X})$ as opposed to the scalar group parameter $\alpha(\vec{X})$ in electrodynamics: $\hat{g}\psi = \exp(i\alpha)\psi$ [19]. That is why the second-rank tensor $A_{ij}$ will be the OP compensating field with local transformational properties of the translations subgroup: $D_j^l \eta_l = \left( \dfrac{\partial}{\partial X_j} - i \sum_p \nu_p A_{pj}^l \right) \eta_l$ (where $\nu_p$ - is the phenomenological charge of the dislocation, it defines the minimal Burgers vector [18]). It is obvious that one second-rank tensor $A_{ij}$, will be enough to compensate all vectors in star $\{\vec{k}\}$. In [18] it was proven that compensating field $A_{ij}$ of the local OP is the distortion tensor. It sets the density of dislocations

$\rho_{pj} = -e_{jkn}(\partial A_{pn}/\partial X_k)$, by definition [20]. Nonequilibrium potential is a functional of the distortion tensor, that gives rise to the vortex structure of equations of state $\delta\Phi/\delta A_{ij} = 0$. Gradient invariance of equations of state in this case is dictated by translational invariance of the local Landau potential.

It is known from the Kadic-Edelen gauge theory of dislocations [21] that there is an analogy between Lorentz force $f_i = e_{ijk} j_j B_k$ (where $\vec{j}$ is vector of current, $B_k = e_{klm} \partial A_m/\partial X_l$ is magnetic induction, $A_m$ is the electromagnetic potential) and Peach-Koehler force $f_i = e_{ijk}\sigma_{nj}\rho_{nk}$ (where $\sigma_{nj}$ is the stress tensor) [22]. In the local Landau theory stress tensor $\sigma_{nj} = \delta\Phi/\delta A_{nj}$ depends on OP components as well as current depends on the wave function in the field theory of electrodynamics $j_m = \delta\Phi/\delta A_m$.

Thus, the inhomogeneous distribution of magnetic moment will be given by a set of functions $\vec{m}_{\vec{k}_l}(\vec{X})$, which constitute coefficients preceding basic functions. In this case the basic functions are obtained from a direct product of the vector representation that characterises three components of magnetic moment and the distribution density of the magnetic moment with the star of vector $\{\vec{k}\}$. The interaction of the external magnetic field and the magnetic OP deforms the magnetic sublattice. When $\vec{k} = \vec{k}(\vec{X})$ inhomogeneous distribution of magnetic moment leads to appearance of inner stresses and dislocations. Since the difference between Peach-Koehler force and Lorentz force in structure is only in contraction by the first index of the stress tensor and dislocation density, then the interaction between magnetic moment and density of dislocations has a structure similar to Abrikosov vortices [12].

Detailed description of magnetic vortex structures is beyond the scope of this paper. However, we can at least note one distinction in the description in connection with the symmetry. Term $\vec{M}[\vec{\nabla}\vec{M}]$ in [15] does not allow for a flat model of vortex solutions that lie, for example, in plane $(xy)$, as distinct from the model with compensating fields. Indeed, if $\vec{M} = (M_1(x,y), M_2(x,y), 0)$ (here we use universally accepted designations $x \equiv X_1$, $y \equiv X_2$), then $[\vec{\nabla}\vec{M}] = (0, 0, \partial M_2/\partial x - \partial M_1/\partial y)$, which results in $\vec{M}[\vec{\nabla}\vec{M}] = 0$.

The interaction between magnetic moment and density of dislocations as well as the interaction between current and magnetic induction has a symmetrical flat solution that depends on coordinate $z$. Indeed, if the distortion tensor and stress tensor lie in plane $(xy)$: $A_{nj}$, $\sigma_{nj}$, where $n, j = 1, 2$, then density of dislocations has the form: $\rho_{n3} = \partial A_{n1}/\partial y - \partial A_{n2}/\partial x$. Therefore, Peach-Koehler force $f_i = e_{ij3}\sigma_{nj}\rho_{n3}$ lies in plane $(xy)$, because $i \neq 3$.

Most likely, magnetic vortex structure of thin films [11] constitutes dislocations of the magnetic sublattice that interact with the magnetic moment. Such an interaction is defined by the distortion tensor in the extended derivative of the magnetic OP in the inhomogeneous potential and is analogous to the phenomenological description of the electron-phonon interaction in the theory of superconductivity [18].

## 6. Phase diagram.

In order to show the phenomenological coefficient region of the potential where solutions obtained in (6) can be correct let us examine section $a_3 > 0$, $b_3 = 0$, $b_4 = 0$. Let us construct a phase diagram of states in the space of parameters $a_1$, $b_1$, $b_2$ assuming that only $a_1 = a_1(T)$ depends on temperature (see Fig.1). Then the domain of existence of the helicoidal phase is limited by condition $\rho_h^2 \geq 0$: $a_1 + a_2 q_h + a_3 q_h^2 \leq 0$. Here $q_h^2$ is defined by the second equation in system (6) and it is satisfied by the plane that goes through axis $a_1$: $b_1 = -q_h^2 b_2$. The surface $a_1 + a_2 q_h + a_3 q_h^2 = 0$ is tangent to the plane $a_1(T_0) = a_2^2/4a_3$ in a straight line $b_1 = -q_0^2 b_2$, where $q_0 = -a_2/2a_3$. It should be noted that plane $a_1(T_0) = a_2^2/4a_3$ is a plane of the second-order transition into inhomogeneous state that is determined by the presence of Lifshitz invariant ($a_2 \neq 0$) in the potential. The domain of existence of homogeneous low-symmetry states is limited by condition $a_1 \leq 0$ (for, $b_1 > 0$, $\eta_1 = 0$, $\eta_2 \neq 0$, for $b_1 < 0$, $\eta_1 = \eta_2 \neq 0$).

Since a general solution for the system of Euler-Lagrange equations (3) is not found, then, in the general case, it is not possible to find out at what values of parameters $a_1$, $b_1$, $b_2$ homogeneous and helicoidal phases satisfy the absolute minimum of the nonequilibrium thermodynamic potential.

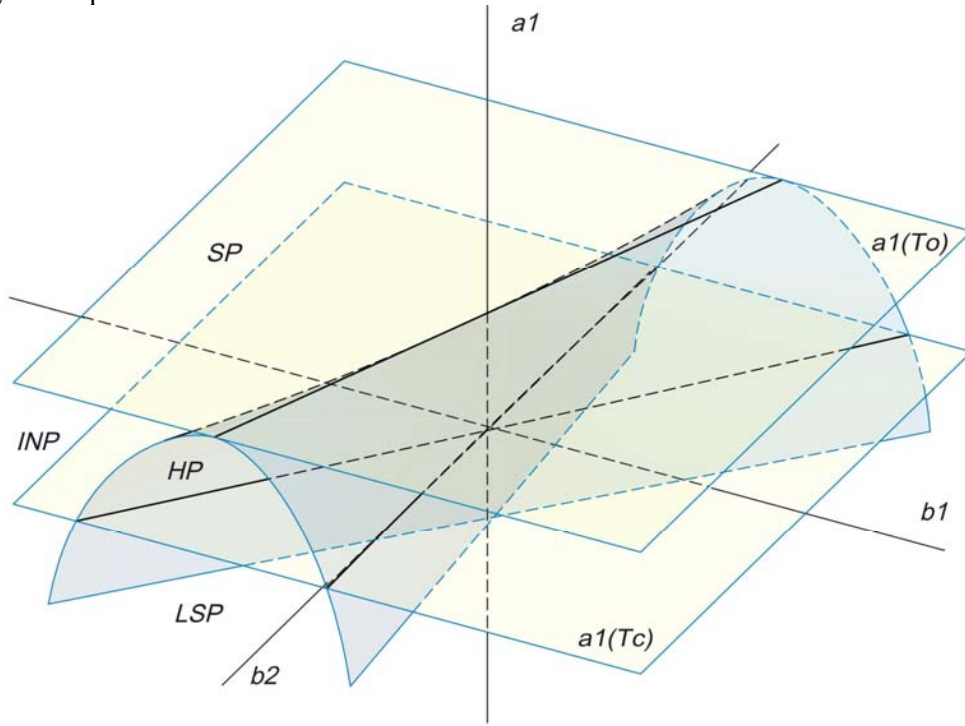

Phase diagram in the neighbourhood of the second-order phase transition point. Phases: SP - high-symmetry, INP - inhomogeneous, HP – helicoidal, LSP - homogeneous low-symmetry.

However, in the small temperature neighbourhood of the second-order phase transition point $a_1(T_0)$ solution $\eta_1 = \rho \cos q_0 z$, $\eta_2 = \rho \sin q_0 z$ satisfies the absolute minimum of Landau potential. Therefore, it can be stated that in the neighbourhood of line

$$a_1 = \frac{a_2^2}{4a_3}, \quad b_1 = -\frac{a_2^2}{4a_3^2}b_2 \tag{11}$$

a helicoidal phase occurs.

Let us examine section $b_2 = 0$ that satisfies the Dzyaloshinskii model. In this case the domain of existence of the helicoidal phase is satisfied by ray $b_1 = 0$, $a_1 \leq a_2^2/4a_3$. This is a known solution [5] that is obtained in anisotropy approximation. When $b_1 \neq 0$ the inhomogeneous phase has a corresponding essentially inhomogeneous solution of the system (3) where the OP module and phase depend on the space coordinate and there are no solutions that would satisfy helicoidal phase.

It should be noted that in any transversal two-dimensional section of a general three-dimensional phase diagram (Fig.1) there is no more than one isolated point in which the helicoidal phase borders on the homogeneous phase. This point is a second-order transition point.

Thus, at decreasing temperature $T \leq T_0$, in the general case, the high-symmetry phase goes into the inhomogeneous phase, different from the helicoidal phase, by means of a second-order transition above Curie point $T_c$ of the possible phase transition to homogeneous low-symmetry phase. Then, at decreasing temperature a helicoidal ordering can occur by means of a first-order transition, because $\Phi_h = 0$ at the border of existence of the helicoidal phase. Similarly, since at the border of existence of the homogeneous low-symmetry phase ($a_1(T_c) = 0$) helicoidal state is satisfied with a greater minimum of the thermodynamic potential as opposed to homogeneous state $\Phi_h \triangleleft \Phi_0 = 0$, then in the general case the formation of the homogeneous low-symmetry phase occurs as a first-order transition. A second-order transition from the high-symmetry phase directly into the helicoidal phase is only possible along line (11) that lies in plane $a_1(T_0) = 0$. On phase diagram $p-T$ this line is represented by a point; here it is assumed that $b_1$ and $b_2$ depend on pressure.

In order to define the point of the second-order phase transition into inhomogeneous state, in [5] the authors expanded the inhomogeneous solution in a Fourier series and found the first harmonic at which energy turned negative. For this purpose it will suffice to insert expression (4) into quadratic approximation of the potential (2) $\Phi' = a_1\rho^2 + a_2 q\rho^2 + a_3 q^2 \rho^2$. This expression takes minimal value when $\partial\Phi'/\partial q = a_2\rho^2 + 2a_3 q\rho^2 = 0$, from which follows $q_0 = -a_2/2a_3$. We used this result when we were defining the plane of the second-order transition $a_1(T_0) = a_2^2/4a_3$. Value $q_0 = -a_2/2a_3$ is the value of the first harmonic Fourier series expansion of an inhomogeneous solution but not a period of the helicoidal phase Such was an assumption initially [5]. However, paper [16] began to associate the period of the helicoid with value

$q_0 = -a_2/2a_3$ ; let us demonstrate that it is not the case. Helicoidal state is satisfied with solution (4). In order to find the period of the helicoid (4) needs to be inserted into the equations of state. For the quadratic approximation of potential $\Phi'$, we obtain $a_1 + a_2 q + a_3 q^2 = 0$ from (6). It is obvious that equation $a_1 + a_2 q + a_3 q^2 = 0$ obtained from the system of equations of state and the minimum condition $a_2 + 2a_3 q = 0$ are not compatible for any coefficients $a_i$. Thus, value $q_0 = -a_2/2a_3$ does not correspond to the period of helicoid but corresponds to the value of the first harmonic Fourier series expansion of an inhomogeneous solution. As follows from the phase diagram on Fig.1, it is into inhomogeneous state that the second-order transition from the high-symmetry phase occurs when the IR allows for Lifshitz invariant: $a_2 \neq 0$. Papers [14, 15] use invariant $\vec{M}[\vec{\nabla} \times \vec{M}]$ as a generalised Lifshitz invariant for crystals without an inversion centre, but this does not alter the essence of the matter.

### 7. Analogy between inhomogeneous helimagnetics and second-order superconductors.

Until now by a helicoidal phase we understood an inhomogeneous long-period structure (4). As is known [23], it is also possible to describe helicoidal structures within a homogeneous IR model with disproportionate vector $\vec{k}$ that does not correspond to the chosen point of the Brillouin zone. Such a description, in essence, postulates the existence of the space periodic distribution of OP, and the IR with disproportionate vector $\vec{k}$ can describe a second-order phase transition.

When describing inhomogeneous deformations in the helimagnetics OP with disproportionate vector $\vec{k}$ that occupies a general position in the Brillouin zone, a problem should be examined when $\eta_l = \eta_l(\vec{X})$, $\vec{k} = \vec{k}(\vec{X})$. This will enable us to describe deformation of helimagnetics for the case of inhomogeneous distortions caused by the external magnetic field. While describing the deformation of helimagnetics in the magnetic field, transition to an inhomogeneous model with $\vec{k} = \vec{k}(\vec{X})$ will be also mostly generic for the inhomogeneous representation $\vec{k} = \vec{k}_0$ examined in this paper. Such a transition is determined by the analogy between the phase diagram for inhomogeneous helimagnetics in the magnetic field [11] and phase diagram of second-order superconductors [12].

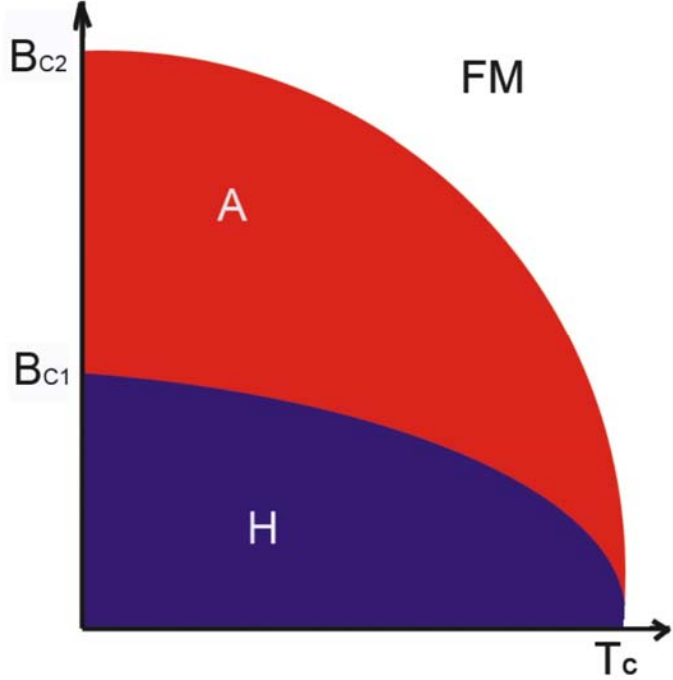

Figure 2. Phase diagram $Fe_{0.5}Co_{0.5}Si$ [11] in the magnetic field: H – helimagnetic, A-phase or "skyrmion" phase, FM – ferromagnetic.

Indeed, Fig. 2 demonstrates the analogy between the helicoidal phase H and the superconducting phase; the A-phase and the mixed superconducting state; the ferromagnetic phase FM and the nonsuperconducting phase [volume 9]. By the A-phase (Fig. 2) we will understand inhomogeneous distribution of the magnetic moment with density of dislocations (Chapter 5) similar to Abrikosov vortices in the second-order superconductor [12]. The density of dislocations is analogous to the magnetic field, it occurs in the OP sublattice due to the change in $\vec{k} = \vec{k}(\vec{X})$. Vector $\vec{k}(\vec{X})$ sets the period of the magnetic sublattice, therefore, incompatibilities will occur at the boundary of two regions with different periods. For example, when changes happen in the direction perpendicular to vector $\vec{k}(\vec{X})$, edge dislocations will emerge [18].

As it was demonstrated in [18], states with $\vec{k} = \vec{k}(\vec{X})$ are described by a pair of variables - OP and compensating field $A_{ij}$ that interact with each other. This interaction is determined by an extended derivative similar to the interaction between the superconducting wave function and the electromagnetic potential in [24]. For OP with $\vec{k} = \vec{k}(\vec{X})$, the compensating field is distortion tensor [18]. Since the compensating field is part of the OP extended derivative, then its physical dimensionality (distortion tensor is dimensionless) is not related to physical dimensionality of the OP itself. Such an interaction is universal and is related to the change of the translational symmetry of the sublattice. The external magnetic field, in the general case, leads to inhomogeneous deformations of helicoidal state with $\vec{k}(\vec{X})$. Thus, we can assume that there is a direct dependence between the external magnetic field and the distortion tensor in the inhomogeneous state of helimagnetics.

For the helicoidal phase and the superconducting phase it is fair to use approximation $\rho = const$ in the magnetic field away from the second-order phase transition point. Condition $\rho = const$, as is known [12], leads to London equations for superconducting potential. Equations of state for the magnetic OP with extended derivative are analogous to Ginzburg-Landau equations of state [24] (if correspondence is set between density of dislocation and magnetic induction, see Chapter 5). Therefore, there is an area on the phase diagram (Fig. 2) where the helicoidal phase H remains stable but dislocations are pushed to the edges of the sample, similar to the magnetic field in the superconductor.

With the increasing magnetic field a mixed state - A-phase (Fig.2) occurs where $\rho = \rho(\vec{X})$. In this state a system of vortices from dislocations is formed, similar to Abrikosov vortices. Density of dislocations is linked to ruptures of the magnetic sublattice in the inhomogeneous state. Due to disproportion of vector $\vec{k}(\vec{X})$, the OP sublattice is not linked to the crystal lattice. Therefore, in this case, dislocations are not related to incompatibilities of the crystal lattice; they show themselves in the form of vortices in the A-phase (Fig.2).

With the further increase in the magnetic field, a critical field $B_{C2}$, can be determined where the inhomogeneous state of the helimagnetic is destroyed and it moves to a ferromagnetic state (Fig. 2). In this case, the critical magnetic field depends on density of dislocations $\rho_{pj} = -e_{jkn}(\partial A_{pn}/\partial X_k)$ that is determined by analogy with the critical magnetic field for the superconducting state [12].

To provide an additional proof for the validity of the suggested description elastic properties of the helimagnetic should be examined in each phase of the diagram on Fig. 2. Due to the presence of dislocations in the helimagnetic hardening should be observed. This hardening is associated with inhomogeneity - with the appearance of additional elastic modules in the potential in the form of coefficients preceding OP extended derivatives where distortion tensor is a linear member. Similar hardening is seen in the superconducting state; it manifests itself in anomalous behaviour of elastic modules at the phase transition to the superconducting state [25].

It should be noted that the compensating field is an independent variable. It interacts with the OP when the translational symmetry of the sublattice changes (not necessarily crystal sublattice). Generally speaking, the crystal lattice does not usually interact minimally with the distortion tensor because it would have to be a very fragile lattice subject to weak interaction that can be also easily restored. An example of a "fragile elastic lattice" is liquid crystals. In 1972, De Gennes noticed that the phase diagram of the deformed SmA is equivalent to the phase diagram of second-order superconductors []. On the basis of this observation he formulated a theory for SmA similar to the Ginzburg-Landau theory. But De Gennes used vector field in the extended derivative; this field cannot compensate for the change in vector of director $\vec{n} = \vec{n}(\vec{X})$ for smectic OP (in [] it was assumed that the distance between SmA layers did not change $d = const$, therefore $\vec{k}(\vec{X}) = \vec{n}(\vec{X})/d$). In [27] a model was built with a tensor compensating field which is free of faults of the De Gennes model [28].

Thus, phase diagram [11] is described within a model with extended derivatives by two variables: OP and compensating field $A_{ij}$, by analogy with the superconducting phase diagram (Fig.2) [12]. The sublattice made of magnetic spins and determined by the weak Dzyaloshinskii-Moriya interaction [29] satisfies two requirements of such description: it is fragile (deforms with the occurrence of dislocations) and is easily restored once the external magnetic field is removed (this demonstrates its elastic properties), similar to liquid crystals. Weak Dzyaloshinskii-Moriya interactions result in the formation of the magnetic sublattice with disproportionate vector $\vec{k}$. In the external field such a lattice is deformed which results in inhomogeneous distortions of not only the OP value, but also of vector $\vec{k} = \vec{k}(\vec{X})$.

The phase diagram on Fig. 2 cannot be described only with magnetic OP, even with the presence of vortex invariant $\vec{M}[\vec{\nabla}\vec{M}]$ [15], because in this case there is no answer to the question why a transition happens from one phase into another.

It is worth noting that for MnSi and FeGe [30] there are phase diagrams $B-T$, analogous to Fig. 2, with the only difference in that in the proximity to zero temperatures with a non-zero magnetic field a conic phase occurs with $|\vec{M}| = const$. This is consistent with the Dzyaloshinskii's assumption [6] that away from the phase transition point approximation $\rho = const$ can be used.

**8. Conclusion.**

Using the IRBI method [9] within an inhomogeneous Lifshitz approach in Landau theory [8] we obtained the exact sinusoidal solution for the system of equations of state that describes helicoidal ordering in crystals. As follows from the second equation of system (6), the presence of the sinusoidal solution is directly linked to anisotropy of the crystal, since the period of the helicoid $q$ is determined only by anisotropic coefficients of the potential. Until now it was thought that anisotropic invariants do not allow obtaining the exact periodic solution corresponding to the helicoidal phase. In the general case, when OP components depend on $\vec{X}$ and the IRBI contains all the space derivatives of the OP components, a sinusoidal solution does exist, in which the phase takes the form $\varphi = \vec{q}\vec{X}$. It follows from the equations of state (3,6) that the existence of the sinusoidal solution does not depend on the presence of the Lifshitz invariant in the nonequilibrium Landau potential. Helicoidal solution for the system of equations of state exists if one takes into account anisotropic gradient invariants in the potential that were obtained using the IRBI method. It becomes obvious now that if the inhomogeneous potential contains anisotropic invariants of the OP components, then anisotropic invariants composed of space derivatives of the OP components must be taken into account in it.

In Chapter 6 we constructed a phase diagram of states in the space of coefficients $a_1(T)$, $b_1$, $b_2$ and obtained the domain of existence for the helicoidal phase. It is important to note that the domain of existence of a stable exact periodic solution on Fig.1 is transversal.

As it is clear from Fig.1, on the phase diagram the domain of existence of inhomogeneous state different from helicoidal state does not coincide with possible helicoidal ordering. It was previously thought that it was one and the same state [5,16], because the section that was examined had all $b_i = 0$, except for $b_1 \neq 0$. But in such conditions a sinusoidal solution for the system of equations of state (6) corresponding to the helicoidal phase does not exist.

We did not expect that the expression for the period of the helicoid obtained from the system of equations of state without approximations: $b_1 + b_2 q^2 + 4b_3 q^3 - 3b_4 q^4 = 0$ is in no way connected with the expression for the period of the helicoid obtained from quadratic approximation for a nonequilibrium potential: $a_1 + a_2 q + a_3 q^2 = 0$. In non-linear Landau theory, the linear approximation cancels the OP module $\rho_h = 0$ (see the first equation of system (6)), therefore, there are no internal contradictions in the theory. It is obvious that linear approximation cannot be used in non-linear Landau theory by definition. However, it would appear that for small values of the OP the period of the helicoid would be close in value to expression $a_1 + a_2 q + a_3 q^2 = 0$ obtained through linear approximation, but this is not the case. Generally speaking, linear approximation has never coincided with the first harmonic $a_2 + 2a_3 q = 0$ expansion into Fourier series of an inhomogeneous solution (see Chapter.6). Nonetheless, the first harmonic of the inhomogeneous solution is still associated with the period of the helicoid, see references to paper [14,16].

Another important result consists in the fact that in the presence of Lifshitz invariant the second-order transition goes into essentially-inhomogeneous state with the corresponding solution $\rho = \rho(z)$, $\partial \varphi / \partial z = f(z)$, and not into helicoid state with the corresponding solution $\eta_1 = \rho \cos qz$, $\eta_2 = \rho \sin qz$ (Fig. 1). The phase transition into the helicoidal phase does not occur, in the general case, through the second-order phase transition. It follows from (6) that the presence of Lifshitz invariant does not affect in principle the existence of the sinusoidal solution. That is why in the general case the transition into the helicoidal phase occurs as a first-order phase transition. This transition is possible from the homogeneous phase in the absence of Lifshitz invariant as well as from the inhomogeneous phase in the case when the IR allows for Lifshitz invariant. This conclusion refers to the formation of long-period helicoidal structures. This conclusion refers to the formation of long-period helicoidal structures described by the OP with vector $\vec{k}_0$ that corresponds to the chosen point of the Brillouin zone. It does not apply to the case when helicoidal structures are described by the OP with disproportionate vector $\vec{k}$ of the general location.

In the case of inhomogeneous deformations of helimagnetics [11,30,31] induced by the magnetic field, one should take account the change in transformation properties of the OP - $\vec{k} = \vec{k}(\vec{X})$. Such description is similar to the description of the superconducting state by Ginzburg-Landau; it results in appearance of an additional compensating field - the distortion tensor that minimally interacts with the OP. Comparative analysis of Fig. 2 demonstrates that the phase diagram of the helimagnetic in the magnetic field is analogous to the phase diagram of the second-order superconductor. It follows from the minimal interaction between the magnetic OP and distortion tensor that the observed vortex states of helimagnetic [11] constitute states with dislocations that are formed due to deformation of the magnetic sublattice and not by the states of skyrmions [15].

**Bibliography**.

[1] H. Boller, A. Kallel Sol. State Comm. 9, 1699, (1971).
[2] M. Iizumi, I.D. Axe , G. Shirane, K. Shimooka, Phys. Rev. B. 15, 4392, (1977).
[3] Y. Ishikawa, M. Arai, J. Phys. Soc. Jpn. 53, 2726 (1984).
[4] Yu.A. Izyumov, Neutron Diffraction on Long-period Structures [in Russian], Energoatomizdat, Moscow (1987).
[5] I.E. Dzyaloshinskii, Zh. Eksp. Teor. Fiz. 46, 1420, (1964).
[6] I.E. Dzyaloshinskii, Zh. Eksp. Teor. Fiz 47, 992, (1964).
[7] V.A. Golovko, Sov. Phys. Solid State 22, 2960, (1980).

[8] E.M. Lifshitz Zh. Eksp. Teor. Fiz. 11, 255 (1941).

[9] Yu.M. Gufan, Sov. Phys. Solid State 13, 225, (1971).

[10] J. C. Toledano and P. Toledano, The Landau Theory of Phase Transitions. World Scientific (1987).
[11] X.Z. Yu, Y. Onose, N. Kanazawa, J.H. Park, J.H. Han, Y. Matsui, N. Nagaosa, Y. Tokura, Nature 465, 901, (2010).
[12] E.M. Lifshitz, L.P. Pitaevskii Statistical Physics, Part 2. Theory of Condensed Matter, Volume 9, Nauka, Moscow (1987).
[13] O.V. Kovalev, Irreducible representations of the space groups. Translated from the Russian, by A. Murray Gross. New York, Gordon and Breach, (1965).
[14] A.N. Bogdanov, D.A. Yablonsky, Zh. Eksp. Teor. Fiz. 95, 178, (1989).
[15] U.K. R¨oßler, A.N. Bogdanov, C. Pfleiderer, Nature 442, 797 (2006).
[16] P. Bak, H.M. Jensen, J. Phys. C: Solid State Phys. 13 L881 (1980).
[17] S.V. Grigoriev, et al., Phys. Rev. B 81, 012408 (2010).
[18] A.Ya. Braginsky, Arxiv preprint, arXiv:1209.0214. (2013).
[19] N.N. Bogolyubov and D.V. Shirkov, Quantum Fields, Nauka, Moscow (1980).
[20] L.D. Landau and E.M. Lifshitz, Theory of Elasticity, Volume 7, Nauka, Moscow (1987).
[21] A. Kadic, D.G. Edelen, A Gauge Theory of Dislocations and Disclinations. Lecture Notes in Physics. Heidelberg: Springer 174, 168 (1983).
[22] M. Lazar, Mathematics and Mechanics of Solids 16, 253 (2011).
[23] Yu.A. Izyumov and V.N. Syromyatnikov, Phase transitions and crystal symmetry. Kluwer, Dordrecht, (1990).
[24] V.L. Ginzburg, L.D. Landau, Zh. Eksp. Teor. Fiz. 20, 1064 (1950).
[25] N.V. Anshukova, et al., Pisma v Zh. Eksp. Teor. Fiz. 46, 373 (1987).
[26] P.G. De Gennes, Solid State Commun. 10, 753 (1972).
[27] A.Ya. Braginsky, Phys. Rev. B67, 174113 (2003).
[28] B.I. Halperin, T.C Lubensky, Solid State Commun. 14, 997 (1974).
[29] I.E. Dzyaloshinskii, J.Phys.Chem.Solids 4, 241 (1958).
[30] S. Mühlbauer, B. Binz, F. Jonietz, C. Pfleiderer, A. Rosch, A. Neubauer, R. Georgii, P. Böni, Science 323, 915, (2009).
[31] X. Z. Yu, N. Kanazawa, Y. Onose, K. Kimoto,W. Z. Zhang, S. Ishiwata, Y. Matsui and Y. Tokura, Nature Materials 10, 106 (2011).